\documentclass{article}
\usepackage[utf8]{inputenc}
\usepackage{multirow}
\usepackage{graphicx}
\usepackage{longtable}
\usepackage{comment}
\usepackage{url}
\usepackage{hyperref}

\usepackage[dvipsnames]{xcolor,colortbl}

\usepackage{soul}
\usepackage{authblk}

\title{Lack of evidence for correlation between COVID-19 infodemic and vaccine acceptance}

\author[1]{Carlo M. Valensise}
\author[2,3]{Matteo Cinelli}
\author[4, 6]{Matthieu Nadini}
\author[5]{Alessandro Galeazzi}
\author[2]{Antonio Peruzzi}
\author[7]{Gabriele Etta}
\author[2]{Fabiana Zollo}
\author[4,6,*]{Andrea Baronchelli}
\author[7,*]{Walter Quattrociocchi}

\affil[1]{Enrico Fermi Research Center, Piazza del Viminale, 1 — 00184 Roma (IT)}
\affil[2]{Ca’ Foscari, University of Venice — Department of Environmental Sciences, Informatics and Statistics, Via Torino 155, 30172 Venezia (IT)}
\affil[3]{Italian National Research Council — Institute for Complex Systems Via dei Taurini 19, 00185 Roma (IT)}
\affil[4]{City University of London,  Department of Mathematics,  London EC1V 0HB, UK}
\affil[5]{University of Brescia, Via Branze, 59 — 25123 Brescia (IT)}
\affil[6]{The Alan Turing Institute, British Library, 96 Euston Road, London NW12DB, UK}
\affil[7]{Sapienza University of Rome — Department of Computer Science, Viale Regina Elena, 295 — 00161 Roma (IT)}
\affil[*]{Corresponding authors: abaronchelli@turing.ac.uk, walter.quattrociocchi@uniroma1.it}

\date{\today}

\begin{document}
\maketitle

\begin{abstract}
How information consumption affects behaviour is an open and widely debated research question.
A popular hypothesis states that the so-called infodemic has a substantial impact on orienting individual decisions~\cite{Loomba2021,Burki2019}. 
A competing hypothesis stresses that exposure to vast amounts of even contradictory information has little effect on personal choices~\cite{pennycook2020fighting, Schmelz2020}.
The COVID-19 pandemic offered an opportunity to investigate this relationship, analysing the interplay between COVID-19 related information circulation and the propensity of users to get vaccinated.
We analyse the vaccine infodemics on Twitter and Facebook by looking at 146M contents produced by 20M accounts between 1 January 2020 and 30 April 2021. 
We find that vaccine-related news triggered huge interest through social media, affecting attention patterns and the modality in which information was spreading. However, we observe that such a tumultuous information landscape translated only in minimal variations in overall vaccine acceptance as measured by Facebook's daily COVID-19 Trends and Impact Survey~\cite{fbsurvey} (previously known as COVID-19 World Symptoms Survey) on a sample of 1.6M users. 
Notably, the observation period includes the European Medicines Agency (EMA) investigations over blood clots cases potentially related to vaccinations, a series of events that could have eroded trust in vaccination campaigns. 
We conclude the paper by investigating the numerical correlation between various infodemics indices and vaccine acceptance, observing strong compatibility with a null model.
This finding supports the hypothesis that altered information consumption patterns are not a reliable predictor of collective behavioural change. Instead, wider attention on social media seems to resolve in polarisation, with the vaccine-prone and the vaccine-hesitant maintaining their positions.
\end{abstract}

\maketitle

\section*{Main}

Social media platforms have radically changed how we access information and are often used as a proxy to understand social mood and opinion trends. Users online tend to acquire information adhering to their beliefs and ignore information dissenting from their views~\cite{bakshy2015exposure, DelVicario2016a, sunstein2017republic}. This process, combined with the unprecedented amount of information available online, has fostered the emergence of groups of like-minded individuals framing and reinforcing a shared narrative (i.e., echo chambers)~\cite{adamic2005political, flaxman2016filter, DelVicario2016, stewart2019information,terren2021echo}. Furthermore, recent studies provided evidence for the effect of feed algorithms in bursting polarisation of social dynamics~\cite{blex2020positive, Cinelli2021}. 
Such a scenario may be considered a fertile environment for misinformation spreading, and an eventual threat for democracies~\cite{vosoughi2018spread,pennycook2021psychology}. 

However, the effect of the interplay between online information diffusion and offline user behaviour is still an open scientific question, with fundamental societal implications especially during a crisis like the ongoing pandemic~\cite{Loomba2021,zarocostas2020fight,cinelli2020covid,yang2021covid}. 
Indeed, the World Health Organisation raised concerns about the effects of the so-called ``infodemic", defined as ``overabundance of information - some accurate and some not - that occurs during an epidemic"~\cite{infodemic_def,tangcharoensathien2020framework}, on global health.
Two main hypotheses compete in accounting for the impact of information consumption on human behaviour.
The first states that infodemic has a substantial impact on orienting individual decisions~\cite{Loomba2021,Burki2019}. The second view stresses that exposure to vast amounts of even contradictory information has little effect on personal choices~\cite{pennycook2020fighting, Schmelz2020}. 

In this work we investigate the possible connection between online information and offline behaviour, by looking at possible correlation between the COVID-19 infodemic and vaccine acceptance as measured by intention to getting a vaccine.
First, we analyse the news diet of 20M unique users on Facebook and Twitter over 16 months (from 1 January 2020 to 30 April 2021) to investigate the relationship between online discussions about COVID-19 vaccines and offline vaccine acceptance rate, in the six European countries (Denmark, France, Germany, Italy, Spain, and United Kingdom) that were mostly involved in the investigation performed by European Medicines Agency (EMA) about cases of blood clots occurred after some vaccinations 
To measure how the online debate may reverberate in offline intentions, we consider Facebook's daily COVID-19 Trends and Impact Survey~\cite{fbsurvey} on a sample of 1.6M users. 
Considering the overall observation period, we find that vaccine announcements triggered users' engagement on social media massively. However, we do not observe significant variations in the vaccine acceptance rate during the same period.Second, to further investigate this lack of evidence for a correlation between infodemic and vaccine acceptance, we focus on the effects of the temporary suspension of the AstraZeneca (now Vaxzevria) vaccine issued by several EU countries and EMA. Also in this case minimal variation in the vaccine acceptance curves was observed after the information related to this event started circulating.
Finally, we extend our analysis to 43 countries worldwide and corroborate our findings by testing correlation between vaccine acceptance curves and several infodemic indices, as measured by the COVID-19 Infodemic observatory~\cite{fbk_obs}.

\subsection*{Evolution of the Vaccine Debate}

\begin{figure}[h!]
  \centering
   \includegraphics[width=0.95\linewidth]{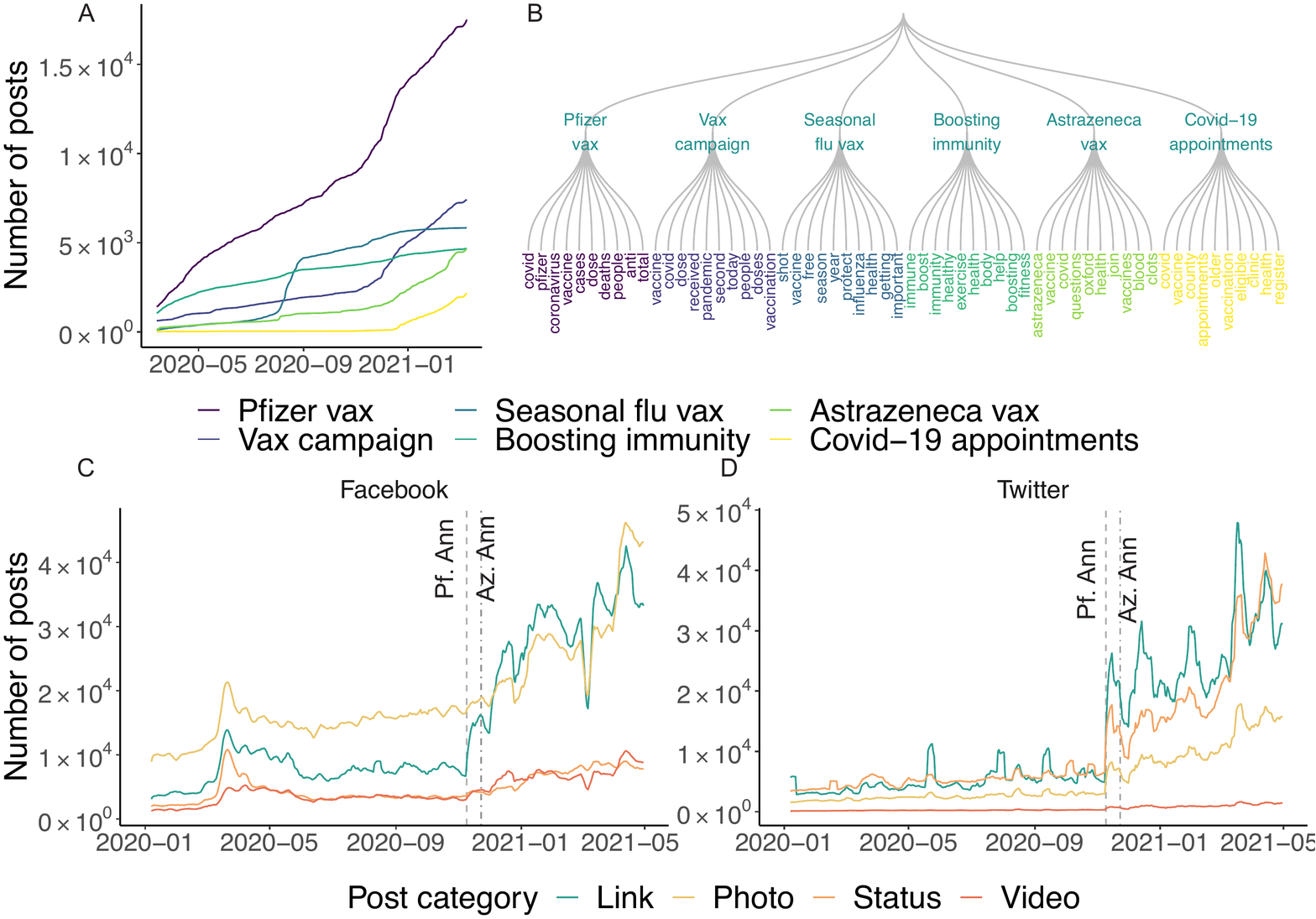}
 \caption{\textbf{Vaccine debate on social media platforms.} Panel A: evolution of debated topics over time. Panel B: Keywords representing the main topics. Panel C and D: seven days moving average of posts divided by category for Facebook and Twitter respectively. Dashed lines represent the announcement of Pfizer and AstraZeneca COVID-19 vaccine effectiveness occurred on 18 November and 23 November 2020, respectively.}
\label{fig:general_stat}
 \end{figure}

We analyse the social media debate around vaccine-related topics on Facebook and Twitter, collecting a large corpus of posts selected via keyword search (see Methods). 
First, we perform topic modelling on the dataset, employing a Deep Learning based approach (see Methods). In this way, we can assign posts to different arguments while studying their evolution over time. Figure~\ref{fig:general_stat} reports the most debated topics.
Consistently with the search, most of the keywords are related to the vaccines and the general vaccination campaign.
``Pfizer vax," ``Vax campaign," ``AstraZeneca Vax" and ``COVID-19 appointments" share a similar trend as they all start increasing around November 2020, which coincides with the disclosure of Pfizer~\cite{pfizer2020announcement} and AstraZeneca~\cite{astrazeneca2020announcement} vaccine efficacy statistics. The debate about ``Seasonal flu vax" peaked around autumn 2020 while ``Boosting immunity" shows a roughly constant growth rate.

\begin{figure}[!htb]
  \centering
   \includegraphics[width=0.95\linewidth]{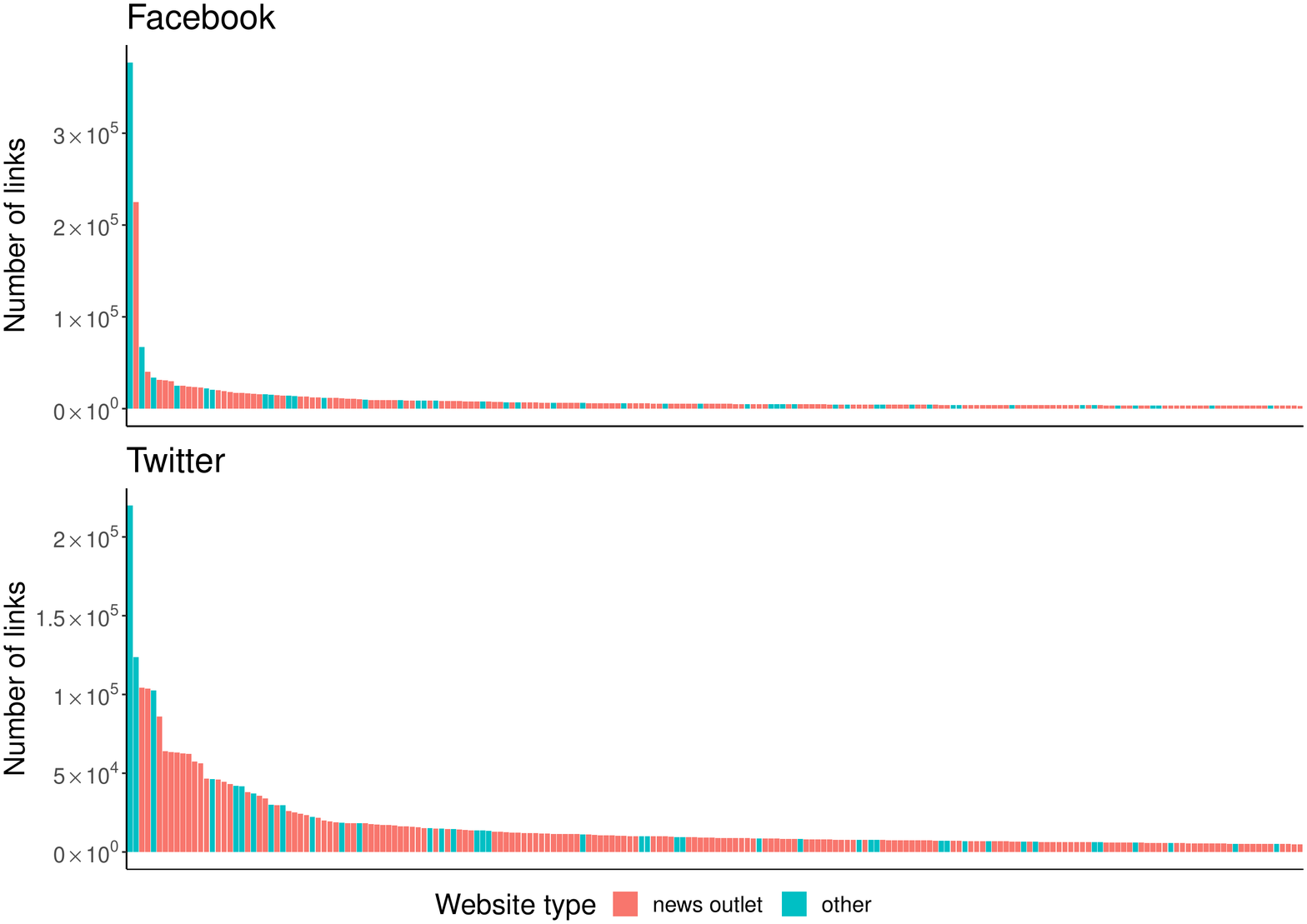}
 \caption{\textbf{Distribution of news outlet sources in the online debate.} Classification of news outlet websites among top 200 domains by number of links for Facebook (top) and Twitter (bottom). All links not pointing to a news source, such as the ones referring to scientific journals or to social network platforms are classified as ``other'': the highest bar in the top figure refers to ``facebook.com'' domain, while in the bottom panel the first and second bars represent  ``youtube.com'' and ``instagram.com'' respectively.} 
\label{fig:news_domain_stat}
 \end{figure}

Next, we consider the social media traffic of different post categories: Link, Photo, Status, Video. In the case of Facebook (Figure~\ref{fig:general_stat}C), we observe an increase in the number of links after the announcement of Pfizer's efficacy on 18 November 2020. This increase unfolded into a surpass of the Link category over Photo after the AstraZeneca efficacy announcement on 23 November 2020, which marked a shift in the most frequent type of content posted. A similar dynamic can be observed on Twitter (Figure~\ref{fig:general_stat}D). 
Since external links usually point to news or scientific articles (see Figure~\ref{fig:news_domain_stat} of SI), we may argue that the increase of links circulation on social media right after vaccines announcement is a signal of a sudden information void that was promptly fulfilled~\cite{tangcharoensathien2020framework}. 
This behaviour may reflect the urge of people to inform themselves about vaccines, whose safety has always been a matter of discussion and one of the main arguments used by the anti-vax community~\cite{johnson2020online}.

\begin{figure}[h!]
  \centering
   \includegraphics[width=0.95\linewidth]{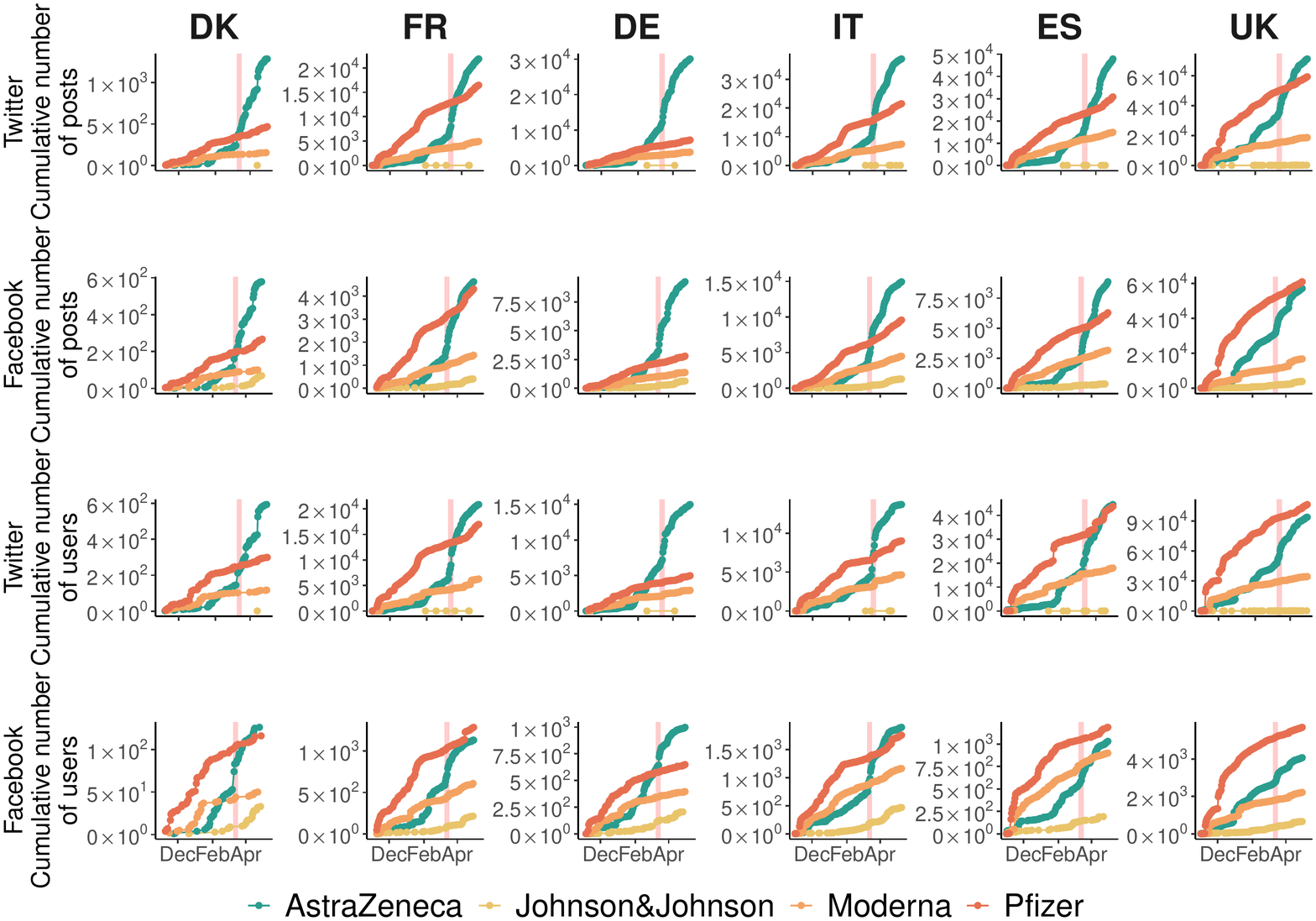}
 \caption{\textbf{Social media debate on COVID-19 vaccines.} Cumulative number of posts and cumulative number of users on Twitter and Facebook by country and vaccine type. Red area represent the period of EMA investigations on AstraZeneca vaccine (from 9 March to 18 March 2021).}
\label{fig:az_stats}
 \end{figure}

We now focus on a specific event of paramount relevance during the general European vaccination. On 10 March 2021, the European Medicines Agency (EMA) reported a rare incidence of blood clots after the vaccine inoculation that induced many European countries to implement a temporal suspension of AstraZeneca~\cite{ema1,ema2}.
The event triggered a relevant increase of traffic on social media, as shown in Figure~\ref{fig:az_stats}. Specifically, we report the temporal evolution of the social debate -- measured through the cumulative number of posts -- for the first four officially approved vaccines (Pfizer, AstraZeneca, Moderna, and Johnson\&Johnson) on Twitter (first row) and Facebook (second row). The social debate around the selected COVID-19 vaccines reaches an increasingly wide audience as shown by the number of unique users involved on Twitter (third row) and Facebook (fourth row). The volume of posts and users sharply increases in correspondence with the suspension date of the AstraZeneca vaccine for every nation and on both platforms. In contrast, the other three types of vaccines show only a moderate growth.

\subsection*{Evolution of Vaccine Acceptance}

\begin{figure}[!htb]
  \centering
\vspace*{-0.5cm}
\hspace*{-0.5cm}
\includegraphics[scale=0.20]{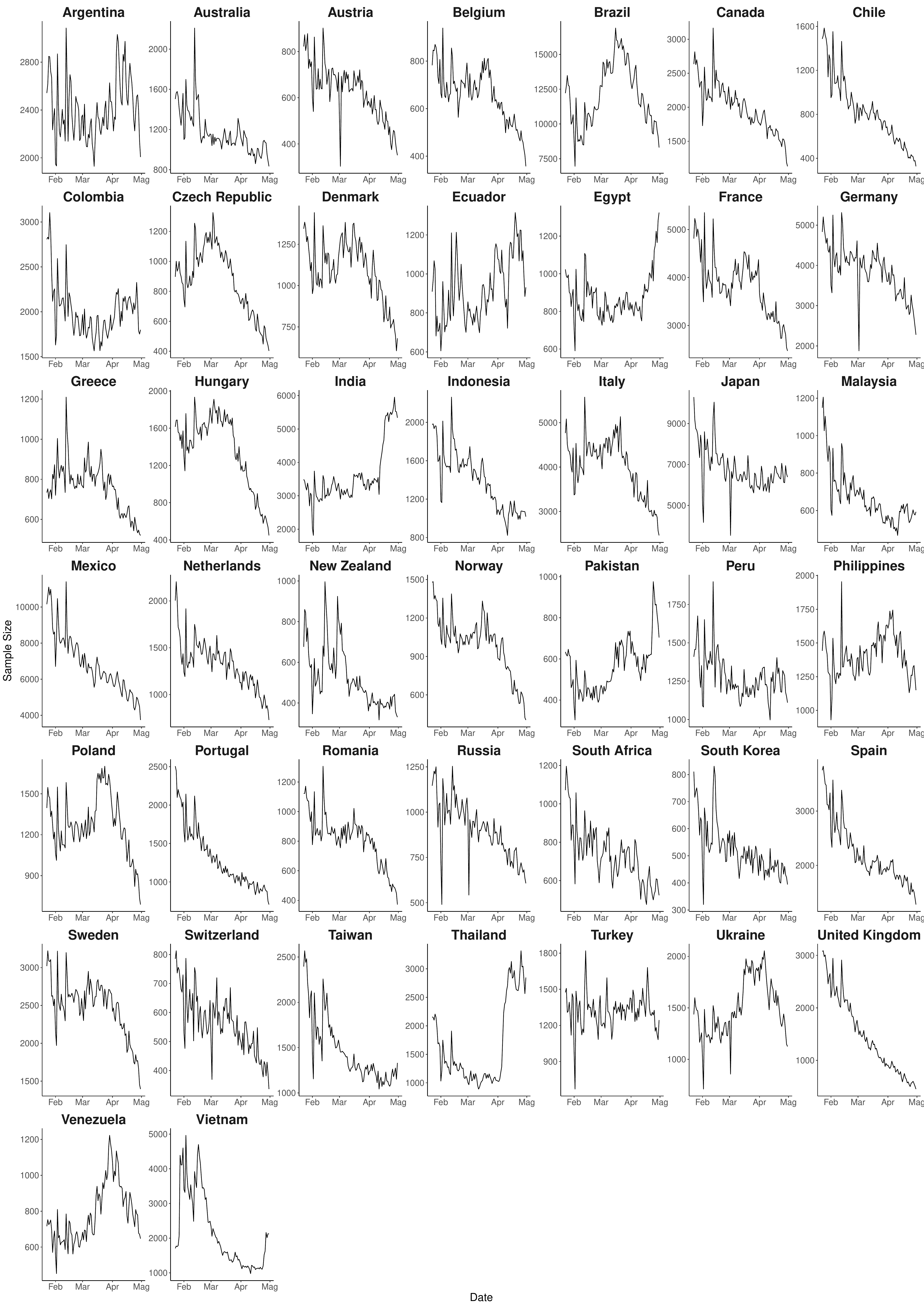}
\caption{\textbf{Daily number of respondents on Facebook Survey.} Solid line: Evolution of the number of respondents for Countries with more than 500 average daily responses. The considered time window ranges from 23 January to 30 April 2021 as for Facebook COVID-19 Trends and Impact Survey.}
\label{fig:countries_respondents}
\end{figure}

We now investigate the effect of these intense online debates on people's actual intentions, as measured by proxies of vaccine acceptance. 
To do so, we leverage data from the COVID-19 Trends and Impact Survey~\cite{fbsurvey,barkay2020weights}, part of the broader Facebook Data for Good project. In this survey, a sample of Facebook users is asked several questions about behaviours and concerns related to COVID-19 on a daily basis. We consider positive answers about vaccination intention from 44 countries (see Figure~\ref{fig:countries_respondents} of SI). The average number of daily respondents is 1954 (10th and 90th percentile 635, 3854, respectively) with a maximum of 12053 (9172, 15196) respondents per day for Brazil and a minimum of 525 (427, 663) respondents per day for South Korea. 
The response to this survey is a valid proxy of the offline users' behaviour~\cite{Zipfel2021,Lessler2021} and a good indicator of vaccine acceptance at the country level.
\begin{figure}[!htb]
  \centering
   \includegraphics[width=0.95\linewidth]{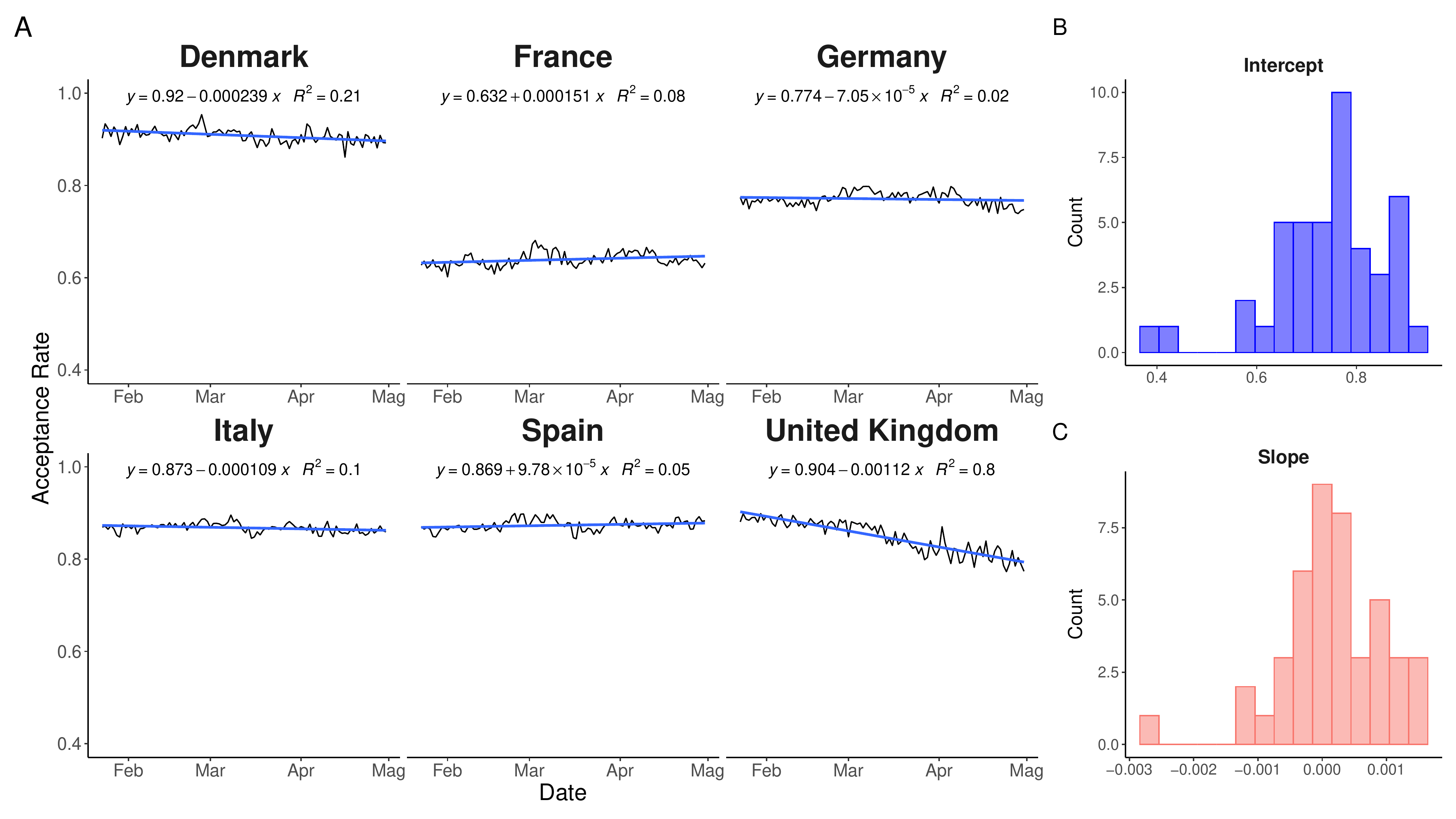}
 \caption{\textbf{Steady acceptance of COVID-19 vaccines.} Panel A: Black lines represent three days moving average of vaccine acceptance rate for Denmark, France, Germany, Italy, Spain, and United Kingdom from 23 January to 30 April 2021 according to Facebook COVID-19 Trends and Impact Survey. The blue lines are the linear fit on the trend. Panel B (C): Histogram of regression intercepts (slopes) for countries with more than 500 average daily respondents.}
 \label{fig:countries6}
 \end{figure}

For each country, we perform a linear fit of the vaccine acceptance rate time series (from 23 January to 30 April 2021), i.e. the percentage of users willing to get vaccinated. Relatively flat trends results for Denmark, France, Germany, Italy, Spain, while United Kingdom shows a slightly decreasing trend (see Figure~\ref{fig:countries6}A). In general, the average slope computed for the whole set of 44 countries (see  Figure~\ref{fig:countries6}C) is $(1.6\pm 8.1) \cdot 10^{-2}$ [\%/day], thus being compatible with close-to-zero values.  
Furthermore, by performing a t-test on the distribution of slopes we fail to reject the (null) hypothesis that the average slope is zero ($t=1.33, p=0.19$).

\begin{figure}[!htb]
  \centering
   \includegraphics[width=0.95\linewidth]{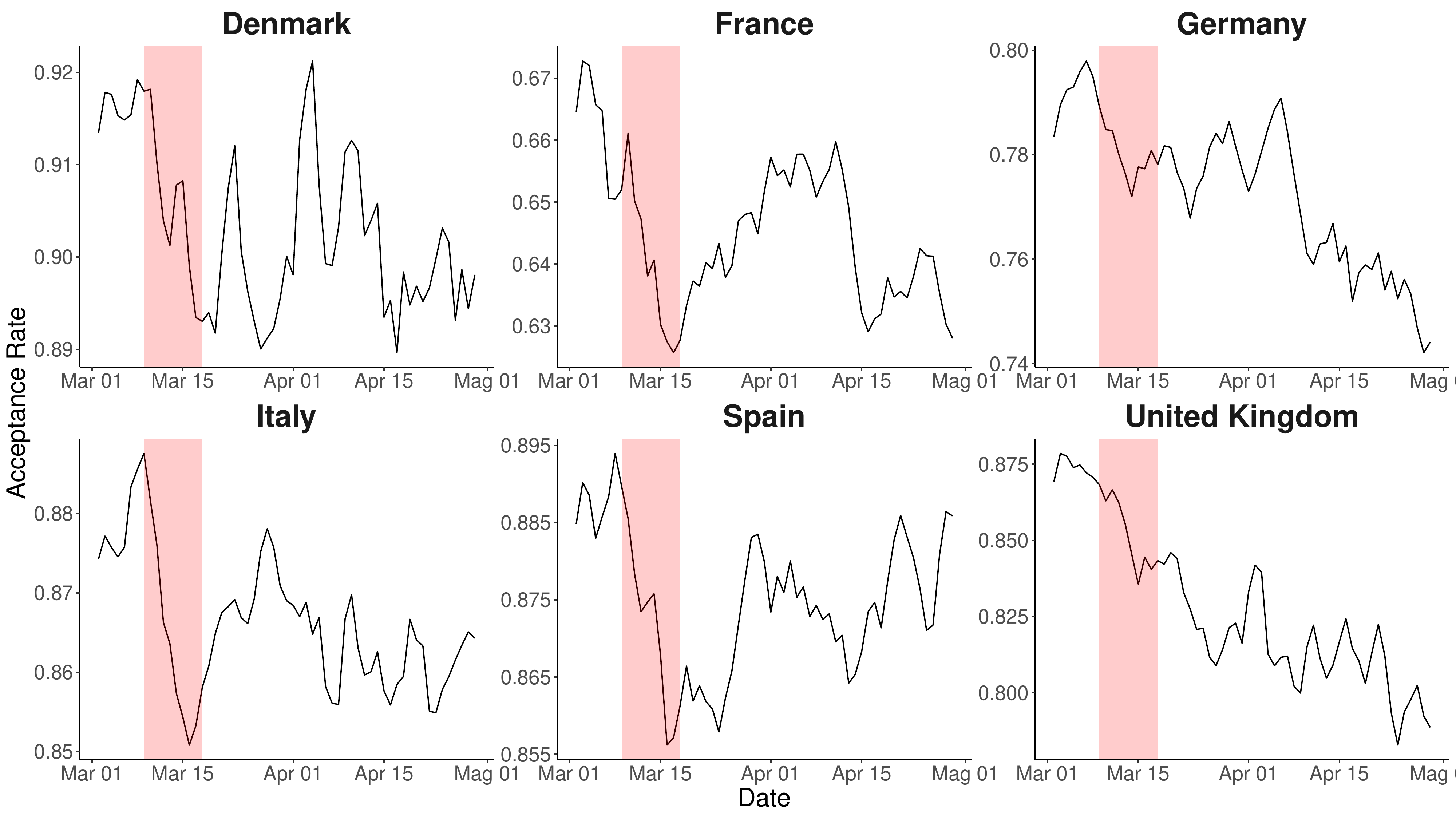}
 \caption{\textbf{Impact of AstraZeneca concerns on vaccine acceptance rate.} Black curve: three days moving average of vaccine acceptance rate for Denmark, France, Germany, Italy, Spain, and United Kingdom from 23 January to 30 April 2021 according to Facebook COVID-19 Trends and Impact Survey. Red area: from the early cases of blood clots (9 March 2021) to the end of EMA investigation (18 March 2021).}
 \label{fig:countries_zoom}
 \end{figure}

In Figure~\ref{fig:countries_zoom} we report a focus on the vaccine acceptance rate in correspondence of the EMA announcement (9 March 2021) about blood clots cases initially occurred in certain EU countries~\cite{ema1}. The \textit{empasse} occurred after 9 March and lasted until 18 March 2021 when EMA declared that the benefits of AstraZeneca vaccination outweigh risks~\cite{ema2}.
In the six considered countries the uncertainty about the side effects of vaccinations translated into a sudden drop of a few percentage points (3-4\%). Thus, despite the great echo generated over the web (see Figure~\ref{fig:az_stats}) by the news about AstraZeneca vaccinations, we do not observe any relevant variation in vaccination intentions.

\subsection*{Correlation between infodemic and vaccine intentions}
To further investigate the possible link between online infodemics and vaccine intentions, we consider the quantitative infodemic indices developed and measured by the COVID-19 Infodemics Observatory \cite{fbk_obs}, and quantify the correlation with vaccine acceptance rates described earlier. 

We consider the Infodemic Risk Index (IRI), the Dynamic Infodemic Risk Index (DIRI), and the number of Tweets. In particular, the IRI is ``estimated indirectly on the basis of the number of followers of users who tweeted, retweeted or quoted unreliable news about COVID-19'', the DIRI is ``estimated directly from users' online endorsement and engagement to evaluate at which rate an user interacts with online messages pointing to potentially unreliable sources of misinformation or disinformation about COVID-19''~\cite{fbk_obs} while the number of Tweets is a simple count.

\begin{figure}[h!]
\centering
\includegraphics[scale = 0.41]{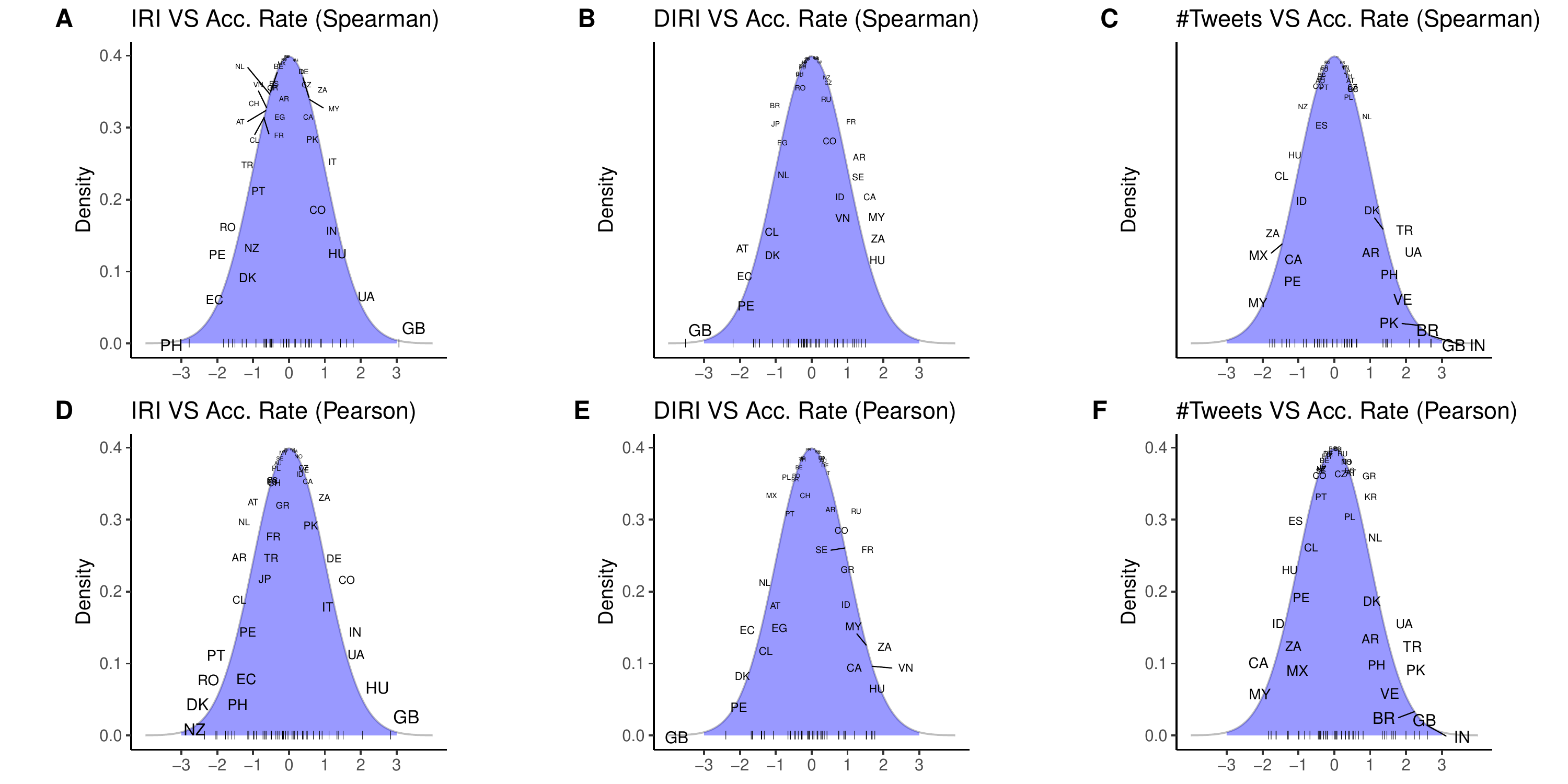}
\caption{Correlation of infodemics indices from the COVID-19 Infodemics Observatory \cite{fbk_obs} and vaccine acceptance rate from Facebook COVID-19 Trends and Impact Survey. As infodemic indices we consider the IRI (panels A and D), the Dyanmic IRI (panels B and E) and the Number of Tweets (panels C and F). The first row displays results obtained by computing the Spearman's correlation coefficient while the second row results are obtained with the Pearson's correlation coefficient. The correlation coefficients on the $x$-axis are reported as standardised values ($z$-scores as explained in Methods). Curves are shaded in blue within 3 standard deviations and the $y$-coordinate of countries (two letter country codes) is represented for visualisation purposes.}
\label{fig:perm_test}
\end{figure}

\begin{figure}
    \centering
    \includegraphics[width=\textwidth]{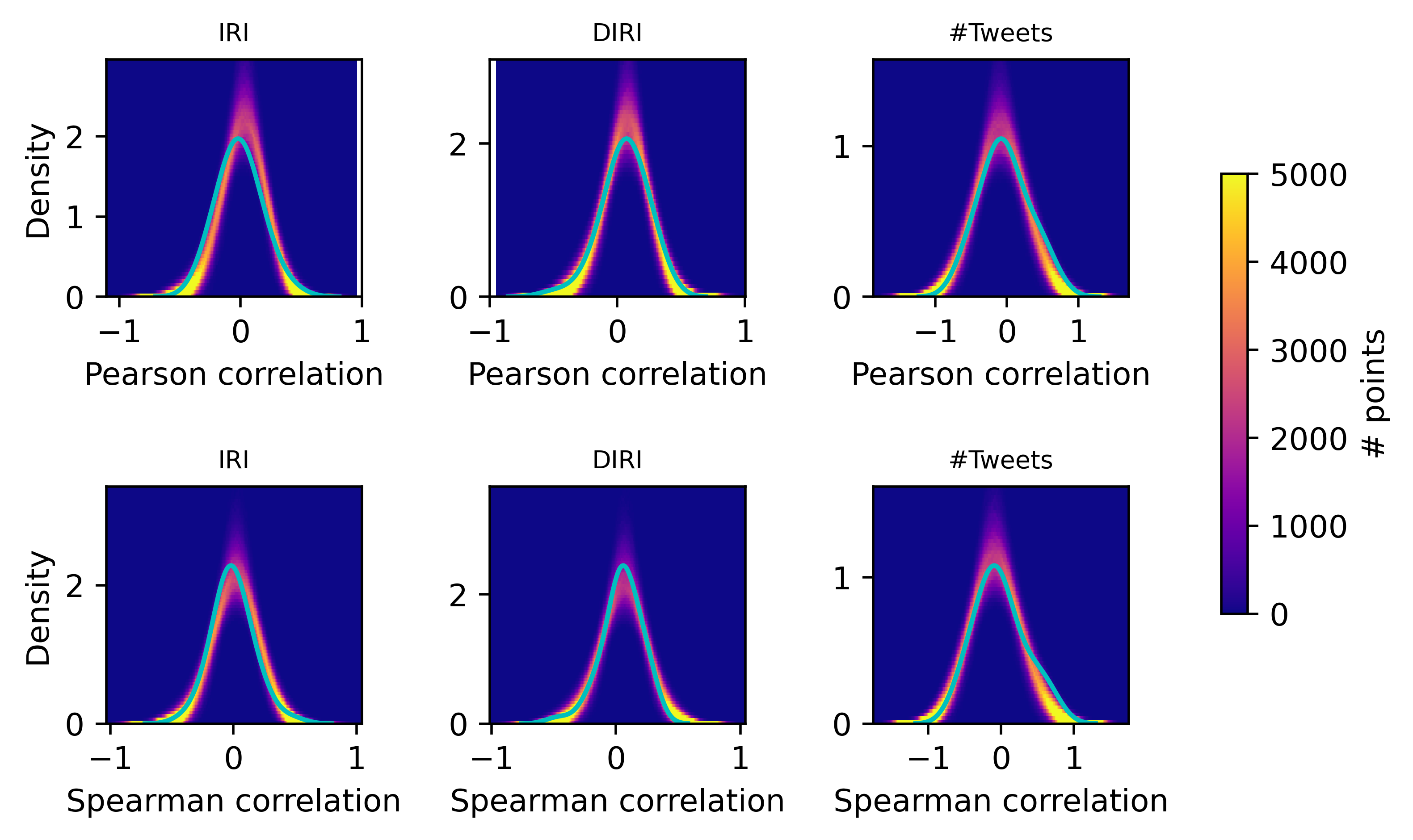}
    \caption{Null model distributions. Each colormap shows the 2d histogram of correlation values computed with 10000 bootstraped samples. In more detail, we repeatedly ($T=10000$) sampled 43 values from the set of correlation values obtained using the null model (see Methods) and computed the corresponding distribution. These 10000 samples from the null distribution are compared with the empirical one, showing that the observed correlation values are indeed compatible with the null model.}
    \label{fig:gaussian_cmap}
\end{figure}

As shown in Figure~\ref{fig:perm_test}, we find that the observed correlations are compatible with a null model in which the relationship between infodemics indices and vaccine acceptance rates is randomised (i.e., the infodemic index of country X is paired with the curve concerning vaccine acceptance of a randomly extracted country Y, and then the correlation between these two signals is measured, see Methods). In fact, we note that (on average) 66\%, 93\% and 99\% of the observed correlations fall within one, two and three standard deviations of the normal distributions deriving from the null model (further details are reported in Table~\ref{tab:sds} and Figure~\ref{fig:gaussian_cmap} of SI). This finding corroborates the hypothesis of a weak or absent link between the exposure to an overabundance of information on social media platforms and offline behaviours.

\subsection*{Discussion}

Our investigation shows that vaccine acceptance has been mostly stable since vaccines were first released, in stark contrast with the major changes in vaccine-related information production and consumption observed on social media platforms over the same period. 

Vaccines for COVID-19 were approved for emergency use. The information landscape became unavoidably turbulent and subject to continuous updates of scientific evidence from many information providers. Great concern was expressed towards the possible impact of misinformation that, leveraging the absence of an established and scientifically shared consensus about vaccines, could limit the effectiveness of vaccination campaigns. More generally, an open research question is whether infodemics might orient people's intention of behaviours. Our results suggest that, at least for COVID-19 vaccinations, this may not be the case, at the aggregated level, and that the evidence is more consistent with the concept of echo chambers enhancing conservatism of opinions. 
This result highlights a crucial methodological aspect, namely that studying the phenomenon only through the lenses of online social media might lead to misleading conclusions. Indeed, the abrupt increase in information consumption does not seem to pair with the almost steady longitudinal behaviour of the acceptance rates.

It is important to highlight the limitations of our study. First, we considered only two social media platforms (Facebook and Twitter) to measure information spreading. As social media platforms widely differ with respect to their structure, audience and content moderation policy, further investigations are required to reveal possibly different correlation scenarios. Second, we considered only some - albeit prominent - definitions of infodemics. Further possibilities exist, and future studies will be able to assess the consistence of our findings across definitions~\cite{simon2021autopsy}. Third, we considered a single survey about vaccine acceptance. However, to the best of our knowledge, the COVID-19 Trends and Impact Survey is the only one that has enough granularity to observe immediate reactions to EMA's investigation. For instance, publicly accessible data from the MIT COVID-19 survey~\cite{MIT-covid-19} provide only 19 data points over nine months while another available survey involves only low and medium income countries~\cite{solis2021covid}. Finally, another potential limitation comes from the timescale of our investigation. Since building cognitive frames and vaccine hesitancy could be slow processes, the analysis of short time periods might be unable to capture long term effects. However, COVID-19 vaccines were announced in late 2020 and by mid 2021 the vaccination campaign was mature in the considered countries, hence short time effects (or lack thereof) are crucial in this case.
On a broader perspective, it is clear that further studies should investigate under which circumstances the link between online information consumption and offline behaviour weakens or strengthens.

\section*{Methods}
\label{sec:methods}
\subsection*{Data Collection}
\label{subsec:data_collection}
Data from Facebook and Twitter were collected covering the period going from 1 January 2020 to 30 April 2021. 

Posts from Facebook were collected using Crowdtangle~\cite{crowdtangle}, a Facebook-owned tool that tracks interactions on public content from Facebook pages and groups, verified profiles, not including paid advertising.
The keyword search was performed by searching for all the possible inflections of the following terms: ``immune", ``dose", ``vaccine", and ``pharma". The search was then extended in order to include vaccines and relative brands. The aforementioned keywords and their inflections were then translated into native languages of the six countries that were taken into account.

Consistently with the queries performed on Crowdtangle, we collected data from Twitter by means of a full-archive historic search within the v2 endpoint and academic research product track. The full list of keywords for the two social media platforms is reported in Tables~\ref{tab:general_terms_by_language} and~\ref{tab:vaccine_keywords_Facebook} of SI.

\subsection*{Topic Modelling}
\label{sec:topic_model_method}
To perform topic modelling we followed the procedure illustrated in~\cite{grootendorst2020bertopic}. For each element of a corpus comprising several posts (including tweets and Facebook posts) we computed an embedding, i.e. a vector representation, through BERT~\cite{devlin2019bert}, a state-of-the-art Natural Language Processing engine. 
After encoding the corpus (with a number $N$ of elements) we obtain a matrix $T$ representing our corpus of size $N\times B$, where $B$ is the dimension of the embedding representation. We leveraged the pre-trained model ``paraphrase-mpnet-base-v2", yielding an embedding of size $B=768$. The encoding of the sentences was obtained through the sBERT package~\cite{reimers-2019-sentence-bert}. Next, to extract the leading topics from the encoded corpus, we applied the HDBSCAN~\cite{McInnes2017} clustering algorithm to the rows of matrix $T$ (after reducing the dimension of the embedding space to 5, via the UMAP algorithm~\cite{mcinnes2020umap}). 
Once all the posts were clustered, we obtained a collection of documents each corresponding to a given topic. To extract the most relevant words, we computed the tf-idf statistics and selected the words with the highest score. The described procedure was applied to 500000 randomly sampled posts, so to fit our computational resources.

\subsection*{Null model}
We performed the following statistical analysis to assess the significance of the correlations observed in real data. For $N=43$ countries (we excluded Taiwan from the analysis since infodemic indicators were unavailable for this country) we consider time series of IRI, DIRI and \#Tweets and the vaccine acceptance rate; each time series spans from 23 January to 30 April 2021. Let us consider the case of IRI for simplicity. First, we compute Spearman correlation coefficient for each country, obtaining an empirical distribution with average $\mu_e$ and standard deviation $\sigma_e$. Next, we compute the correlation for all possible country pairs, obtaining $N(N-1)=1806$ correlation values. For instance, while in the empirical case the IRI of Denmark is compared against the vaccine acceptance rate of Denmark, in the randomisation process we compare the IRI of Denmark with that of all the other countries except Denmark and we iterate this comparison over all countries. We obtain a null gaussian-like distribution parametrised by $\mu_n$, $\sigma_n$. Next, we standardise correlation values $x$ (both randomised and empirical) computing
\[
z = \frac {x-\mu_n}{\sigma_n}\,.
\]
We also repeat the same procedure computing the Pearson correlation coefficient between the three infodemic indicators and the vaccine acceptance rate.
The distribution of $z$ is reported in Figure~\ref{fig:perm_test}.

\section*{Acknowledgements}
The authors acknowledge the 100683EPID Project “Global Health Security Academic Research Coalition” SCH-00001-3391.
Walter Quattrociocchi wants to thank Michele Secci for advice and inspiration.

\bibliographystyle{unsrt}
\bibliography{references.bib}

\begin{thebibliography}{10}

\bibitem{Loomba2021}
Sahil Loomba, Alexandre de~Figueiredo, Simon~J. Piatek, Kristen de~Graaf, and
  Heidi~J. Larson.
\newblock Measuring the impact of {COVID}-19 vaccine misinformation on
  vaccination intent in the {UK} and {USA}.
\newblock {\em Nature Human Behaviour}, 5(3):337--348, 2021.

\bibitem{Burki2019}
Talha Burki.
\newblock Vaccine misinformation and social media.
\newblock {\em The Lancet Digital Health}, 1(6):e258--e259, 2019.

\bibitem{pennycook2020fighting}
Gordon Pennycook, Jonathon McPhetres, Yunhao Zhang, Jackson~G Lu, and David~G
  Rand.
\newblock Fighting {COVID}-19 misinformation on social media: Experimental
  evidence for a scalable accuracy-nudge intervention.
\newblock {\em Psychological Science}, 31(7):770--780, 2020.

\bibitem{Schmelz2020}
Katrin Schmelz.
\newblock Enforcement may crowd out voluntary support for {COVID}-19 policies,
  especially where trust in government is weak and in a liberal society.
\newblock {\em Proceedings of the National Academy of Sciences},
  118(1):e2016385118, 2020.

\bibitem{fbsurvey}
Junchuan Fan, Yao Li, Kathleen Stewart, Anil~R. Kommareddy, Andres Garcia,
  Jinyi Ma, Zheng Liu, Joe O’Brien, Adrianne Bradford, Xiaoyi Deng, Samantha
  Chiu, Frauke Kreuter, Neta Barkay, Alyssa Bilinski, Brian Kim, Tal Galili,
  Daniel Haimovich, Sarah LaRocca, Stanley Presser, Katherine Morris, Joshua~A
  Salomon, Elizabeth~A. Stuart, Ryan Tibshirani, Tali~Alterman Barash, Curtiss
  Cobb, Andi Gros, Ahmed Isa, Alex Kaess, Faisal Karim, Roee Eliat, Ofir~Eretz
  Kedosha, Shelly Matskel, Roee Melamed, Amey Patankar, Irit Rutenberg, Tal
  Salmona, and David Vannette.
\newblock {COVID}-19 world symptom survey data {API}., 2020.
\newblock Carnegie Mellon University, Pennsylvania, United States.

\bibitem{bakshy2015exposure}
Eytan Bakshy, Solomon Messing, and Lada~A Adamic.
\newblock Exposure to ideologically diverse news and opinion on {F}acebook.
\newblock {\em Science}, 348(6239):1130--1132, 2015.

\bibitem{DelVicario2016a}
Michela~Del Vicario, Alessandro Bessi, Fabiana Zollo, Fabio Petroni, Antonio
  Scala, Guido Caldarelli, H.~Eugene Stanley, and Walter Quattrociocchi.
\newblock The spreading of misinformation online.
\newblock {\em Proceedings of the National Academy of Sciences},
  113(3):554--559, 2016.

\bibitem{sunstein2017republic}
Cass~R Sunstein.
\newblock {\em \#Republic}.
\newblock Princeton University Press, 2017.

\bibitem{adamic2005political}
Lada~A Adamic and Natalie Glance.
\newblock The political blogosphere and the 2004 {US} election: divided they
  blog.
\newblock In {\em Proceedings of the 3rd International Workshop on Link
  Discovery}, pages 36--43, 2005.

\bibitem{flaxman2016filter}
Seth Flaxman, Sharad Goel, and Justin~M Rao.
\newblock Filter bubbles, echo chambers, and online news consumption.
\newblock {\em Public Opinion Quarterly}, 80(S1):298--320, 2016.

\bibitem{DelVicario2016}
Michela~Del Vicario, Gianna Vivaldo, Alessandro Bessi, Fabiana Zollo, Antonio
  Scala, Guido Caldarelli, and Walter Quattrociocchi.
\newblock Echo chambers: Emotional contagion and group polarization on
  {F}acebook.
\newblock {\em Scientific Reports}, 6(1), 2016.

\bibitem{stewart2019information}
Alexander~J Stewart, Mohsen Mosleh, Marina Diakonova, Antonio~A Arechar,
  David~G Rand, and Joshua~B Plotkin.
\newblock Information gerrymandering and undemocratic decisions.
\newblock {\em Nature}, 573(7772):117--121, 2019.

\bibitem{terren2021echo}
Ludovic Terren and Rosa Borge-Bravo.
\newblock Echo chambers on social media: A systematic review of the literature.
\newblock {\em Review of Communication Research}, 9:99--118, 2021.

\bibitem{blex2020positive}
Chris Blex and Taha Yasseri.
\newblock Positive algorithmic bias cannot stop fragmentation in homophilic
  networks.
\newblock {\em The Journal of Mathematical Sociology}, pages 1--18, 2020.

\bibitem{Cinelli2021}
Matteo Cinelli, Gianmarco De~Francisci Morales, Alessandro Galeazzi, Walter
  Quattrociocchi, and Michele Starnini.
\newblock The echo chamber effect on social media.
\newblock {\em Proceedings of the National Academy of Sciences},
  118(9):e2023301118, 2021.

\bibitem{vosoughi2018spread}
Soroush Vosoughi, Deb Roy, and Sinan Aral.
\newblock The spread of true and false news online.
\newblock {\em Science}, 359(6380):1146--1151, 2018.

\bibitem{pennycook2021psychology}
Gordon Pennycook and David~G Rand.
\newblock The psychology of fake news.
\newblock {\em Trends in Cognitive Sciences}, 2021.

\bibitem{zarocostas2020fight}
John Zarocostas.
\newblock How to fight an infodemic.
\newblock {\em The Lancet}, 395(10225):676, 2020.

\bibitem{cinelli2020covid}
Matteo Cinelli, Walter Quattrociocchi, Alessandro Galeazzi, Carlo~Michele
  Valensise, Emanuele Brugnoli, Ana~Lucia Schmidt, Paola Zola, Fabiana Zollo,
  and Antonio Scala.
\newblock The covid-19 social media infodemic.
\newblock {\em Scientific Reports}, 10(1):1--10, 2020.

\bibitem{yang2021covid}
Kai-Cheng Yang, Francesco Pierri, Pik-Mai Hui, David Axelrod, Christopher
  Torres-Lugo, John Bryden, and Filippo Menczer.
\newblock The covid-19 infodemic: Twitter versus facebook.
\newblock {\em Big Data \& Society}, 8(1):20539517211013861, 2021.

\bibitem{infodemic_def}
\url{https://www.who.int/teams/risk-communication/infodemic-management}.

\bibitem{tangcharoensathien2020framework}
Viroj Tangcharoensathien, Neville Calleja, Tim Nguyen, Tina Purnat, Marcelo
  D’Agostino, Sebastian Garcia-Saiso, Mark Landry, Arash Rashidian, Clayton
  Hamilton, Abdelhalim AbdAllah, et~al.
\newblock Framework for managing the {COVID}-19 infodemic: methods and results
  of an online, crowdsourced who technical consultation.
\newblock {\em Journal of Medical Internet Research}, 22(6):e19659, 2020.

\bibitem{fbk_obs}
The covid-19 infodemics observatory.
\newblock \url{https://covid19obs.fbk.eu/#/}.

\bibitem{pfizer2020announcement}
Pfizer Team.
\newblock Pfizer and {B}iontech conclude phase 3 study of {COVID}-19 vaccine
  candidate, meeting all primary efficacy endpoints.
\newblock
  \url{https://www.pfizer.com/news/press-release/press-release-detail/pfizer-and-biontech-conclude-phase-3-study-covid-19-vaccine},
  2020.
\newblock Pfizer and BioNTech, United States.

\bibitem{astrazeneca2020announcement}
{AZD}1222 vaccine met primary efficacy endpoint in preventing {COVID}-19.
\newblock
  \url{https://www.astrazeneca.com/media-centre/press-releases/2020/azd1222hlr.html},
  2020.
\newblock AstraZeneca, London, England.

\bibitem{johnson2020online}
Neil~F Johnson, Nicolas Vel{\'a}squez, Nicholas~Johnson Restrepo, Rhys Leahy,
  Nicholas Gabriel, Sara El~Oud, Minzhang Zheng, Pedro Manrique, Stefan Wuchty,
  and Yonatan Lupu.
\newblock The online competition between pro-and anti-vaccination views.
\newblock {\em Nature}, 582(7811):230--233, 2020.

\bibitem{ema1}
{COVID}-19 vaccine {A}strazeneca: {PRAC} preliminary view suggests no specific
  issue with batch used in {A}ustria.
\newblock
  \url{https://www.ema.europa.eu/en/news/covid-19-vaccine-astrazeneca-prac-preliminary-view-suggests-no-specific-issue-batch-used-austria},
  2021.
\newblock EMA, Amsterdam, The Netherlands.

\bibitem{ema2}
{COVID}-19 vaccine {A}strazeneca: {PRAC} investigating cases of thromboembolic
  events - vaccine’s benefits currently still outweigh risks.
\newblock
  https://www.ema.europa.eu/en/news/covid-19-vaccine-astrazeneca-prac-investigating-cases-thromboembolic-events-vaccines-benefits,
  2021.
\newblock EMA, Amsterdam, The Netherlands.

\bibitem{barkay2020weights}
Neta Barkay, Curtiss Cobb, Roee Eilat, Tal Galili, Daniel Haimovich, Sarah
  LaRocca, Katherine Morris, and Tal Sarig.
\newblock Weights and methodology brief for the {COVID}-19 symptom survey by
  {U}niversity of {M}aryland and {C}arnegie {M}ellon {U}niversity, in
  partnership with {F}acebook.
\newblock {\em arXiv preprint arXiv:2009.14675}, 2020.

\bibitem{Zipfel2021}
Casey~M. Zipfel, Vittoria Colizza, and Shweta Bansal.
\newblock The missing season: The impacts of the {COVID}-19 pandemic on
  influenza.
\newblock {\em Vaccine}, 39(28):3645--3648, 2021.

\bibitem{Lessler2021}
Justin Lessler, M.~Kate Grabowski, Kyra~H. Grantz, Elena Badillo-Goicoechea,
  C.~Jessica~E. Metcalf, Carly Lupton-Smith, Andrew~S. Azman, and Elizabeth~A.
  Stuart.
\newblock Household {COVID}-19 risk and in-person schooling.
\newblock {\em Science}, 372(6546):1092--1097, 2021.

\bibitem{simon2021autopsy}
Felix~M Simon and Chico~Q Camargo.
\newblock Autopsy of a metaphor: The origins, use and blind spots of the
  ‘infodemic’.
\newblock {\em New Media \& Society}, page 14614448211031908, 2021.

\bibitem{MIT-covid-19}
Collis A., Garimella K., Rahimian~M.A. Moehring~A., Babalola S., Gobat N.,
  Shattuck D., J.~Stolow, Eckles D., and Aral S.
\newblock Global survey on covid-19 beliefs, behaviors, and norms.
\newblock {\em Technical report, MIT Sloan School of Management}, 2020.

\bibitem{solis2021covid}
Julio~S Sol{\'\i}s~Arce, Shana~S Warren, Niccol{\`o}~F Meriggi, Alexandra
  Scacco, Nina McMurry, Maarten Voors, Georgiy Syunyaev, Amyn~Abdul Malik,
  Samya Aboutajdine, Opeyemi Adeojo, et~al.
\newblock Covid-19 vaccine acceptance and hesitancy in low-and middle-income
  countries.
\newblock {\em Nature medicine}, 27(8):1385--1394, 2021.

\bibitem{crowdtangle}
CrowdTangle Team.
\newblock Crowdtangle.
\newblock 2020.
\newblock Facebook, Menlo Park, California, United States.

\bibitem{grootendorst2020bertopic}
Maarten Grootendorst.
\newblock {BERT}opic: Leveraging {BERT} and {c-TF-IDF} to create easily
  interpretable topics.
\newblock \url{https://github.com/MaartenGr/BERTopic}, 2020.

\bibitem{devlin2019bert}
Jacob Devlin, Ming-Wei Chang, Kenton Lee, and Kristina Toutanova.
\newblock {BERT}: {P}re-training of deep bidirectional transformers for
  language understanding.
\newblock {\em arXiv preprint arXiv:1810.04805}, 2019.

\bibitem{reimers-2019-sentence-bert}
Nils Reimers and Iryna Gurevych.
\newblock Sentence-{BERT}: Sentence embeddings using siamese {BERT}-networks.
\newblock In {\em Proceedings of the 2019 Conference on Empirical Methods in
  Natural Language Processing}. Association for Computational Linguistics,
  2019.

\bibitem{McInnes2017}
Leland McInnes, John Healy, and Steve Astels.
\newblock hdbscan: Hierarchical density based clustering.
\newblock {\em The Journal of Open Source Software}, 2(11), 2017.

\bibitem{mcinnes2020umap}
Leland McInnes, John Healy, and James Melville.
\newblock {UMAP}: Uniform manifold approximation and projection for dimension
  reduction.
\newblock {\em arXiv preprint arXiv:1802.03426}, 2020.

\end{thebibliography}

\newpage

\newpage
\cleardoublepage
\section*{Supplementary Information}

\begin{table}[h!]
\footnotesize
\begin{tabular}{|l|l|l|}
\hline
\textbf{Country}  & \textbf{Vaccine keywords}   & \textbf{Brand Keywords}                          \\ \hline
\textbf{Denmark} & \begin{tabular}[c]{@{}l@{}}immun*, dosis, vaccin*, medic*, \\ apotek, no-vax*, novax*, pro-vax*,  \\ provax*, antivax*, anti-vax* \end{tabular}  & 
\multirow{6}{*}{\begin{tabular}[c]{@{}l@{}} \vspace{0.16cm} AstraZeneca, \\\vspace{0.16cm} BioNTech, \\\vspace{0.16cm} CanSino,\\\vspace{0.16cm} CureVac, \\\vspace{0.16cm} Jassen, \\\vspace{0.16cm} Johnson \& Johnson,  \\\vspace{0.16cm} Moderna, \\\vspace{0.16cm} Novavax, \\\vspace{0.16cm} Pfizer, \\\vspace{0.16cm} Sinopharm, \\\vspace{0.16cm} Sputnik \end{tabular}} \\ 
\cline{1-2}
\textbf{France} & \begin{tabular}[c]{@{}l@{}}immun*, dos*, pharma*,  \\ médicament*, no-vax*, novax*, \\pro-vax*, provax*, antivax*, anti-vax*\end{tabular}  &    \\ 
\cline{1-2}
\textbf{Germany} & \begin{tabular}[c]{@{}l@{}}immun*, pharma*, vakz*,  \\ impf*, no-vax*, novax*, pro-vax*,  \\ provax*, antivax*, anti-vax*\end{tabular}   &        \\ \cline{1-2}
\textbf{Italy} & \begin{tabular}[c]{@{}l@{}}immun*, farma*, pharma*, dos*,  \\ vacc*, no-vax*, novax*, pro-vax*, \\ provax*, antivax*, anti-vax*\end{tabular} & \\   \cline{1-2}
\textbf{Spain} & \begin{tabular}[c]{@{}l@{}}inmun*, vacuna*, dosi*, farma*,  \\ pharma*, no-vax*,  novax*, pro-vax*, \\ provax*, antivax*, anti-vax*\end{tabular} & \\  \cline{1-2}
\textbf{United Kingdom}  & \begin{tabular}[c]{@{}l@{}}vaccin*, dose*, pharma*, immun*,  \\ no-vax*, novax*, pro-vax*, provax*,  \\ antivax*, anti-vax*\end{tabular}   &        \\ \hline
\end{tabular}
\caption{\textbf{Vaccine-related keywords.} List of vaccine-related terms used for the 6 EU countries under consideration to collect data both from Twitter and Facebook. The terms ending with an asterisk indicate that they were expanded to obtain a set of meaningful terms sharing the same root. The extended list of Facebook keywords is visible in Table~\ref{tab:vaccine_keywords_Facebook}.}
\label{tab:general_terms_by_language}
\end{table}

\begin{table}[h!]
\footnotesize
\begin{tabular}{|l|l|}
\hline
                 & \textbf{Facebook}                         \\ \hline
\textbf{Astrazeneca}      & \begin{tabular}[c]{@{}l@{}}Astrazeneca, astrazeneca, AstraZeneca, AZD1222, ChAdOx1, \\ ChAdOx1 nCoV-2019, Vaxzevria\end{tabular}                                               \\ \hline
\textbf{Moderna} & mRNA-1273, moderna, Moderna                                                                                                                                                    \\ \hline
\textbf{Pfizer}  & \begin{tabular}[c]{@{}l@{}}pfizer, biontech, pfizer/BionTech, BNT16b2, \\ Pfizer, Pfizer/BionTech, pfizer/biontech\end{tabular}                                                \\ \hline
\textbf{J\&J}    & \begin{tabular}[c]{@{}l@{}}JNJ-78436735, Johnson \& Johnson, J\&J, j\&j, j and j, \\ johnson and johnson, Johnson and Johnson, Janssen\end{tabular}                            \\ \hline
\textbf{Sputnik} & \begin{tabular}[c]{@{}l@{}}sputnik, Sputnik V, sputnik v, Gam-COVID-Vac, \\ Gam-COVID-Vac Sputnik V, Sputnik, \\ Gamaleya, Gamaleya Scientific Research Institute\end{tabular} \\ \hline
\end{tabular}
\caption{\textbf{Extended keywords for Facebook.} List of vaccine brand terms employed in the collection of all posts from Facebook.}
\label{tab:vaccine_keywords_Facebook}
\end{table}

 \begin{figure}[!htb]
  \centering
\vspace*{-0.5cm}
\hspace*{-0.5cm}
   \includegraphics[scale=0.20]{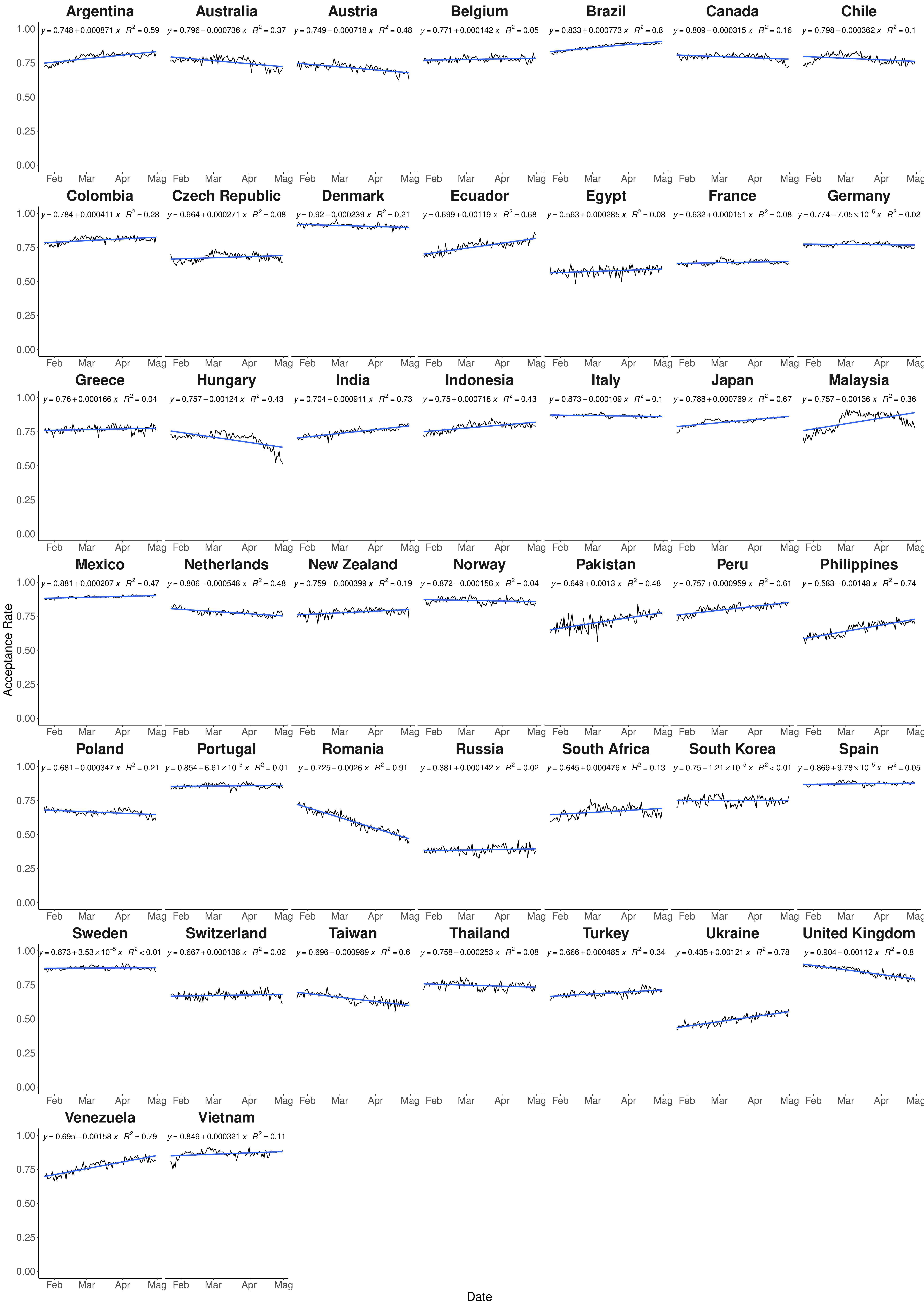}
 \caption{\textbf{Acceptance rates of COVID-19 vaccines.} Black lines represent three days moving average of vaccine acceptance rate for Countries with more than 500 average daily responses on average from 23 January to 30 April 2021 as for Facebook COVID-19 Trends and Impact Survey. Blue lines are linear fit on the trend.}
 \label{fig:countries_acceptance}
\end{figure}

\begin{table}[h!]
\center
 \resizebox{.98\textwidth}{!}{\begin{tabular}{lcccccccc}
\hline  \hline
\textbf{Country}        & \textbf{SampleSize} & \textbf{Daily Avg. Sample Size} & \textbf{Sd Sample Size} & \textbf{Median} & \textbf{Intercept} & \textbf{Slope}              & \textbf{(sd)} & \textbf{(p-value)} \\ \hline
\textbf{Argentina}      & 236533              & 2413.602                        & 253.5509                & 0.802           & 0.748              & 0.00087                     & 0.00007       & 0                  \\
\textbf{Australia}      & 114623              & 1169.622                        & 214.7617                & 0.769           & 0.796              & -0.00074                    & 0.0001        & 0                  \\
\textbf{Austria}        & 61308               & 624.3724                        & 122.0722                & 0.716           & 0.749              & -0.00072                    & 0.00008       & 0                  \\
\textbf{Belgium}        & 64553               & 659.3214                        & 113.5003                & 0.777           & 0.771              & 0.00014                     & 0.00007       & 0.03299            \\
\textbf{Brazil}         & 1181246             & 12053.53                        & 2249.532                & 0.882           & 0.833              & 0.00077                     & 0.00004       & 0                  \\
\textbf{Canada}         & 195665              & 1996.582                        & 361.9683                & 0.797           & 0.809              & -0.00031                    & 0.00007       & 0.00003            \\
\textbf{Chile}          & 81302               & 829.6122                        & 301.6523                & 0.781           & 0.798              & -0.00036                    & 0.00011       & 0.00126            \\
\textbf{Colombia}       & 197117              & 2011.398                        & 296.741                 & 0.808           & 0.784              & 0.00041                     & 0.00007       & 0                  \\
\textbf{Czech Republic} & 85395               & 874.1122                        & 224.2357                & 0.68            & 0.664              & 0.00027                     & 0.00009       & 0.00438            \\
\textbf{Denmark}        & 106951              & 1092.735                        & 183.234                 & 0.908           & 0.92               & -0.00024                    & 0.00005       & 0                  \\
\textbf{Ecuador}        & 90684               & 925.3469                        & 163.4877                & 0.768           & 0.699              & 0.00119                     & 0.00008       & 0                  \\
\textbf{Egypt}          & 85953               & 877.0714                        & 123.646                 & 0.58            & 0.563              & 0.00028                     & 0.0001        & 0.00444            \\
\textbf{France}         & 376513              & 3839.378                        & 616.8985                & 0.637           & 0.632              & 0.00015                     & 0.00005       & 0.0049             \\
\textbf{Germany}        & 378931              & 3860.291                        & 631.6381                & 0.772           & 0.774              & -0.00007                    & 0.00005       & 0.17489            \\
\textbf{Greece}         & 74656               & 761.8214                        & 121.8735                & 0.768           & 0.76               & 0.00017                     & 0.00008       & 0.03889            \\
\textbf{Hungary}        & 133041              & 1361.408                        & 389.105                 & 0.713           & 0.757              & -0.00124                    & 0.00015       & 0                  \\
\textbf{India}          & 352456              & 3596.49                         & 862.6574                & 0.754           & 0.704              & 0.00091                     & 0.00006       & 0                  \\
\textbf{Indonesia}      & 136795              & 1395.867                        & 307.4061                & 0.791           & 0.75               & 0.00072                     & 0.00008       & 0                  \\
\textbf{Italy}          & 392368              & 4006.092                        & 643.2691                & 0.869           & 0.873              & -0.00011                    & 0.00003       & 0.00184            \\
\textbf{Japan}          & 665633              & 6792.173                        & 1051.388                & 0.831           & 0.788              & 0.00077                     & 0.00005       & 0                  \\
\textbf{Malaysia}       & 65897               & 672.4184                        & 145.6112                & 0.849           & 0.757              & 0.00136                     & 0.00019       & 0                  \\
\textbf{Mexico}         & 672716              & 6864.449                        & 1650.531                & 0.892           & 0.881              & 0.00021                     & 0.00002       & 0                  \\
\textbf{Netherlands}    & 129335              & 1320.367                        & 259.6571                & 0.778           & 0.806              & -0.00055                    & 0.00006       & 0                  \\
\textbf{New Zealand}    & 51517               & 525.6837                        & 146.8149                & 0.783           & 0.759              & 0.0004                      & 0.00008       & 0.00001            \\
\textbf{Norway}         & 97071               & 990.898                         & 229.0561                & 0.866           & 0.872              & -0.00016                    & 0.00007       & 0.03644            \\
\textbf{Pakistan}       & 55149               & 562.7449                        & 127.2885                & 0.723           & 0.649              & 0.0013                      & 0.00014       & 0                  \\
\textbf{Peru}           & 125468              & 1280.286                        & 139.5034                & 0.811           & 0.757              & 0.00096                     & 0.00008       & 0                  \\
\textbf{Philippines}    & 137937              & 1407.52                         & 160.547                 & 0.67            & 0.583              & 0.00148                     & 0.00009       & 0                  \\
\textbf{Poland}         & 124824              & 1273.306                        & 219.1616                & 0.665           & 0.681              & -0.00035                    & 0.00007       & 0                  \\
\textbf{Portugal}       & 128440              & 1310.883                        & 404.9535                & 0.857           & 0.854              & 0.00007                     & 0.00006       & 0.242              \\
\textbf{Romania}        & 79672               & 812.7857                        & 174.6214                & 0.592           & 0.725              & -0.0026                     & 0.00008       & 0                  \\
\textbf{Russia}         & 88773               & 905.8469                        & 159.4271                & 0.387           & 0.381              & 0.00014                     & 0.00009       & 0.12256            \\
\textbf{South Africa}   & 71884               & 733.5102                        & 147.2221                & 0.674           & 0.645              & 0.00048                     & 0.00013       & 0.00033            \\
\textbf{South Korea}    & 51499               & 525.5                           & 97.51164                & 0.754           & 0.75               & -0.00001                    & 0.0001        & 0.90549            \\
\textbf{Spain}          & 218042              & 2224.699                        & 567.5856                & 0.873           & 0.869              & 0.0001                      & 0.00004       & 0.02061            \\
\textbf{Sweden}         & 242548              & 2474.684                        & 376.1167                & 0.874           & 0.873              & 0.00004                     & 0.00004       & 0.4292             \\
\textbf{Switzerland}    & 55158               & 562.25                          & 103.1644                & 0.678           & 0.667              & 0.00014                     & 0.00009       & 0.12402            \\
\textbf{Taiwan}         & 144199              & 1471.418                        & 363.1374                & 0.651           & 0.696              & -0.00099                    & 0.00008       & 0                  \\
\textbf{Thailand}       & 161474              & 1647.694                        & 724.6496                & 0.748           & 0.758              & -0.00025                    & 0.00009       & 0.00583            \\
\textbf{Turkey}         & 129194              & 1318.306                        & 145.3587                & 0.69            & 0.666              & 0.00048                     & 0.00007       & 0                  \\
\textbf{Ukraine}        & 147594              & 1506.061                        & 277.8563                & 0.498           & 0.435              & 0.00121                     & 0.00007       & 0                  \\
\textbf{United Kingdom} & 144973              & 1480.821                        & 738.6411                & 0.858           & 0.904              & -0.00112                    & 0.00006       & 0                  \\
\textbf{Venezuela}      & 76974               & 785.449                         & 162.0155                & 0.790           & 0.695              & \multicolumn{1}{c}{0.00158} & 0.00008       & 0                  \\
\textbf{Vietnam}        & 213950              & 2183.163                        & 1083.191                & 0.870           & 0.849              & \multicolumn{1}{c}{0.00032} & 0.00009       & 0.00072            \\ \hline \hline
\end{tabular} }

\caption{\textbf{Summary statistics of linear fits. }
}
\label{tab:RegCoeff}
\end{table}

\begin{table}[ht]
\centering
\begin{tabular}{r|ccc|ccc|}
  \hline
  & & Spearman & &  & Pearson & \\  \hline
$\sigma$ & IRI & DIRI & \#Tweets & IRI & DIRI & \#Tweets \\ \hline
  1 & 72.1 & 69.8 & 60.5 & 65.1 & 67.4 & 62.8 \\  
  2 & 95.3 & 95.3 & 88.4 & 88.4 & 95.3 & 93.0 \\ 
  3 & 97.7 & 97.7 & 100.0 & 100.0 & 97.7 & 100.0 \\
   \hline
\end{tabular}
\caption{Percentage of observed correlation values within 1, 2 and 3 standard deviations. The expected percentages of values drawn from a normal distribution within one, two and three standard deviations are 68\%, 95\% and 99.7\% respectively.}
\label{tab:sds}
\end{table}

\end{document}